\def\spose#1{\hbox to 0pt{#1\hss}}
\def\approxlt{\mathrel{\spose{\lower 3pt\hbox{$\sim$}}
        \raise 2.0pt\hbox{$<$}}}
\def\approxgt{\mathrel{\spose{\lower 3pt\hbox{$\sim$}}
        \raise 2.0pt\hbox{$>$}}}

\def\multleft#1{\hbox to size{\vbox {\halign {\lft{##}\cr #1}}\hfill}\par}
\def\multright#1{\hbox to size{\vbox {\halign {\rt{##}\cr #1}}\hfill}\par}

\def\degmark{^\circ}
\def\boxit#1{\vbox{\hrule\hbox{\vrule\kern3pt\vbox{\kern3pt
          #1 \kern3pt}\kern3pt\vrule}\hrule}}

\def\cm{{\rm\thinspace cm}}

\def\K{{\rm\thinspace K}}
\def\keV{{\rm\thinspace keV}}

\def\s{{\rm\thinspace s}}

\def\cmps{\hbox{$\cm\s^{-1}\,$}}

\def\pcmsq{\hbox{$\cm^{-2}\,$}}

\documentclass[12pt,preprint]{aastex}
\usepackage{natbib}
\received{}
\accepted{}

\slugcomment{Scheduled for The Astrophysical Journal v601 n1 Jan 20 2004}

\shorttitle{Chandra X-ray Observations of NGC 4258}

\shortauthors{Young \& Wilson}

\begin{document}

\title{Chandra X-ray Observations of NGC 4258: Iron Absorption Lines from the
  Nucleus}

\author{A. J. Young\altaffilmark{2} and A. S. Wilson\altaffilmark{2}}

\affil{Astronomy Department, University of Maryland, College Park, MD
	20742}

% Notice that each of these authors has alternate affiliations, which
% are identified by the \altaffilmark after each name.  The actual alternate
% affiliation information is typeset in footnotes at the bottom of the
% first page, and the text itself is specified in \altaffiltext commands.
% There is a separate \altaffiltext for each alternate affiliation
% indicated above.

\altaffiltext{1}{Present address MIT Center for Space Research, 77
  Massachusetts Avenue, Cambridge, MA 02139}

\altaffiltext{2}{Adjunct Astronomer, Space Telescope Science
	Institute, 3700 San Martin Drive, Baltimore, MD 21218}

% The abstract environment prints out the receipt and acceptance dates
% if they are relevant for the journal style.  For the aasms style, they
% will print out as horizontal rules for the editorial staff to type
% on, so long as the author does not include \received and \accepted
% commands.  This should not be done, since \received and \accepted dates
% are not known to the author.

\begin{abstract}

We report sub-arcsecond resolution X-ray imaging spectroscopy of the low
luminosity active galactic nucleus of NGC 4258 and its immediate surroundings
with the Chandra X-ray Observatory. NGC 4258 was observed four times, with the
first two observations separated by one month, followed over a year later by
two consecutive observations. The spectrum of the nucleus is well described by
a heavily absorbed ($N_{\rm H} \simeq 7 \times 10^{22} \pcmsq$, which did not
change), hard X-ray power law of variable luminosity, plus a constant, thermal
soft X-ray component.  We do not detect an iron K$\alpha$ emission line with
the upper limit to the equivalent width of a narrow, neutral iron line ranging
between 94 and 887 eV (90\%\ confidence) for the different observations. 
During the second observation on 2000-04-17, two narrow absorption features are
seen with $>$ 99.5\%\ confidence at $\simeq 6.4$ keV and $\simeq 6.9$ keV, which
we identify as resonant absorption lines of Fe XVIII -- Fe XIX K$\alpha$ and Fe
XXVI K$\alpha$, respectively.  In addition, the 6.9 keV absorption line is
probably variable on a timescale of $\sim 6000$ sec. The absorption lines are
analyzed through a curve of growth analysis, which allows the relationship
between ionic column and kinematic temperature or velocity dispersion to be
obtained for the observed equivalent widths.  We discuss the properties of the
absorbing gas for both photo and collisionally ionized models.  Given that the
maser disk is viewed at an inclination $i = 82\degmark$, the gas responsible
for the 6.9~keV absorption line may be in an inner disk, a disk-wind boundary
layer or be thermal gas entrained at the base of the jet. The gas which gives
rise to the photoelectric absorption may be the same as that which causes the
6.4~keV Fe K$\alpha$ absorption provided that the gas has a bulk velocity
dispersion of a few thousand km s$^{-1}$.  This is the first detection of iron
X-ray absorption lines in an extragalactic source with a nearly edge-on
accretion disk, and this phenomenon is likely to be related to similar X-ray
absorption lines in Galactic X-ray binaries with nearly edge-on accretion
disks.

\end{abstract}

% The different journals have different requirements for keywords.  The
% keywords.apj file, found on aas.org in the pubs/aastex-misc directory, 
% contains a list of keywords used with the ApJ and Letters.  These are 
% usually assigned by the editor, but authors may include them in their
% manuscripts if they wish. 


% That's it for the front matter.  On to the main body of the paper.
% We'll only put in tutorial remarks at the beginning of each section
% so you can see entire sections together.

% In the first two sections, you should notice the use of the LaTeX \cite
% command to identify citations.  The citations are tied to the
% reference list via symbolic KEYs.  We have chosen the first three
% characters of the first author's name plus the last two numeral of the
% year of publication.  The corresponding reference has a \bibitem
% command in the reference list below.
%
% Please see the AASTeX manual for a more complete discussion on how to make
% \cite-\bibitem work for you.   

\section{Introduction} \label{sec:intro}

The Seyfert 1.9 galaxy NGC 4258 (M 106) is a prime candidate for the study of
low luminosity active galactic nuclei. The mass of the central black hole in
NGC 4258 is well determined from position and velocity measurements of the
H$_2$O maser emission which show a thin, slightly warped, accurately-Keplerian
disk of radius 0.12 pc to 0.25 pc in orbit around a $( 3.9 \pm 0.1 ) \times
10^7 M_\odot$ black hole \citep{1993Natur.361...45N, 1995ApJ...440..619G,
1995Natur.373..127M, 1999Natur.400..539H}. \citet{1995ApJ...455L..13W} have
found strongly polarized, broad optical emission lines indicative of an
obscured active nucleus. Early X-ray observations with Einstein
\citep{1992ApJ...390..365C, 1992ApJS...80..531F} and ROSAT
\citep{1994A&A...284..386P, 1999A&A...352...64V, 1995ApJ...440..181C} detected
soft, extended X-ray emission, but could not penetrate the gas obscuring the
nucleus. ASCA observations were the first to detect the active nucleus, which
exhibited a power law spectrum of photon index $\Gamma \simeq 1.8$ and an
absorption corrected 2 -- 10 keV luminosity of $L_{\rm x} = 4 \times 10^{40}$
erg s$^{-1}$ obscured by a column density of $N_{\rm H} \simeq 1.5 \times
10^{23}$ cm$^{-2}$ \citep{1994PASJ...46L..77M}. The X-ray luminosity is a small
fraction of the Eddington luminosity, $L_{\rm X} / L_{\rm Edd} \approxlt 3
\times 10^{-5}$. Subsequent ASCA observations have shown the nuclear luminosity
to be variable on year-long timescales with $L_{\rm x}$ in the range $(0.4 - 1)
\times 10^{41}$ erg s$^{-1}$ \citep{2000ApJ...540..143R, 2002ApJS..139....1T,
1999ApJS..120..179P} while the photon index and column density have remained
approximately constant. BeppoSAX observations revealed 100\% continuum
variability on a timescale of a ${\rm few} \times 10^5$ s, and 10 -- 20\%
variability on timescales as short as one hour \citep{2001ApJ...556..150F}.
Interesting variability has also been seen by ASCA in the 6.4 -- 6.9 keV iron
lines. In 1993, a narrow Fe line was seen at $6.66^{+0.20}_{-0.07}$ keV with an
equivalent width (EW) of $180^{+73}_{-64}$ eV \citep{2002ApJS..139....1T},
while in 1996, the Fe line energy had changed to $6.31^{+0.09}_{-0.10}$ keV
with an EW of $54^{+25}_{-27}$ eV \citep{2002ApJS..139....1T}.  This change in
line energy, by 0.35 keV, is significantly larger than the systematic
uncertainty in the ASCA calibration \citep{Iwasawa}.  In a 1999 ASCA
observation, a narrow iron line was observed at $6.45^{+0.10}_{-0.07}$ keV with
an EW of $107^{+42}_{-37}$ eV \citep{2000ApJ...540..143R}. An XMM-Newton
observation in 2000 constrained the EW of a 6.45 keV iron line to be $< 45$ eV
\citep{2002A&A...384..793P}, indicating that the iron line had weakened
significantly since the 1999 ASCA observation (a $\sim 100$ eV EW narrow
iron line would have easily been detected in the XMM-Newton observation).

We have obtained a series of Chandra observations to study various aspects of
NGC 4258. At 7.2~Mpc \citep{1999Natur.400..539H} $1\arcsec = 35$~pc, and we
take the Galactic hydrogen column towards NGC 4258 to be $N_{\rm H} = 1.19
\times 10^{20}$~cm$^{-2}$ \citep{1996ApJS..105..369M}.

\section{Observations and Reduction}

NGC 4258 was observed by Chandra on four separate occasions with ACIS-S
(Table~\ref{tab:data}). The first observation on 2000-03-08 had a duration of
2.9 ksec, with a 128 row sub-array and alternating 0.4~sec and 0.1~sec
frame-times, in order to obtain an approximate count rate and assess the
effects of pile-up. The second observation on 2000-04-17 had a duration of
13~ksec with a full frame and 3.2~sec frame-time, primarily to study the
``anomalous arms'' of NGC 4258 \citep[hereafter WYC]{2001ApJ...560..689W}; the
nucleus suffers from the effects of pile-up. The third observation on
2001-05-28 had a duration of 20~ksec with a full frame and 3.2~sec frame-time,
again to study the anomalous arms, and the nucleus again suffers from the
effects of pile-up. The fourth observation, taken immediately after the third,
on 2001-05-29 had a duration of 7~ksec with a 128 row sub-array and a 0.4~sec
frame-time to mitigate the effects of pile-up and to study the nucleus of NGC
4258.

The latest gain files were applied to the data and background flares were
filtered out by excluding data from those times during which the background
(i.e. source-free regions on the S3 chip) count rate was $> 3\sigma$ above or
below the mean background count rate.  For the 2.9~ksec duration observation,
we used a dead-time correction (DTCOR) of 0.552 to calculate the exposure time
of the 0.4~sec frames. The resulting exposure times are listed in
Table~\ref{tab:data}. We shall refer to the 4 observations as the 1.6 ksec, 13
ksec, 20 ksec and 7 ksec exposures respectively. Note that because the 0.4 sec
frame-time 1.6 ksec exposure was taken in an alternating mode, the 1.6 ksec of
exposure time is spread over the 2.9 ksec duration of the observation. CIAO
2.2.1 with CALDB 2.17 were used, and the spectra were analyzed using XSPEC 11.2
\citep{1996adass...5...17A}. To obtain spectra of the nucleus, counts were
extracted from circular apertures of radius $1\farcs5$ centered on the
brightest pixel, and background counts were extracted from an annulus of inner
radius $4\arcsec$ and outer radius $10\arcsec$ also centered on the nucleus.
The spectra were grouped to $\ge 15$ counts per bin and the $\chi^2$ statistic
was used to compare models to the data.  The confidence region for a model
parameter was determined by varying that parameter until the $\chi^2$ value,
minimized by allowing all of the remaining free parameters to vary, was equal
to the best fit value plus $\Delta \chi^2$ where $\Delta \chi^2 = 2.706$, 6.635
and 7.879 for the 90\%\, 99\%\ and 99.5\%\ confidence regions for a single
interesting parameter, respectively.  Unless stated otherwise 90\%\ confidence
regions have been used.

When analyzing spectra of the nucleus taken with a 3.2~sec frame-time it is
important to correct for the effects of pile-up (the 0.4~sec frame-time
observations do not require pile-up correction), and we used the pile-up
correction routine of \citet{2001ApJ...562..575D} that is incorporated into the XSPEC software
package. We find the grade morphing parameter $\alpha$ (the fraction of ``good
grades'' given by $p$ piled-up photons is $\propto \alpha^{(p-1)}$) to be
$\alpha = 0.41^{+0.06}_{-0.07}$ for the 20 ksec data set and $\alpha =
0.43^{+0.03}_{-0.03}$ for the 13 ksec data set.

\section{X-ray Morphology}

\begin{figure}
\centerline{
  \includegraphics[scale=0.45]{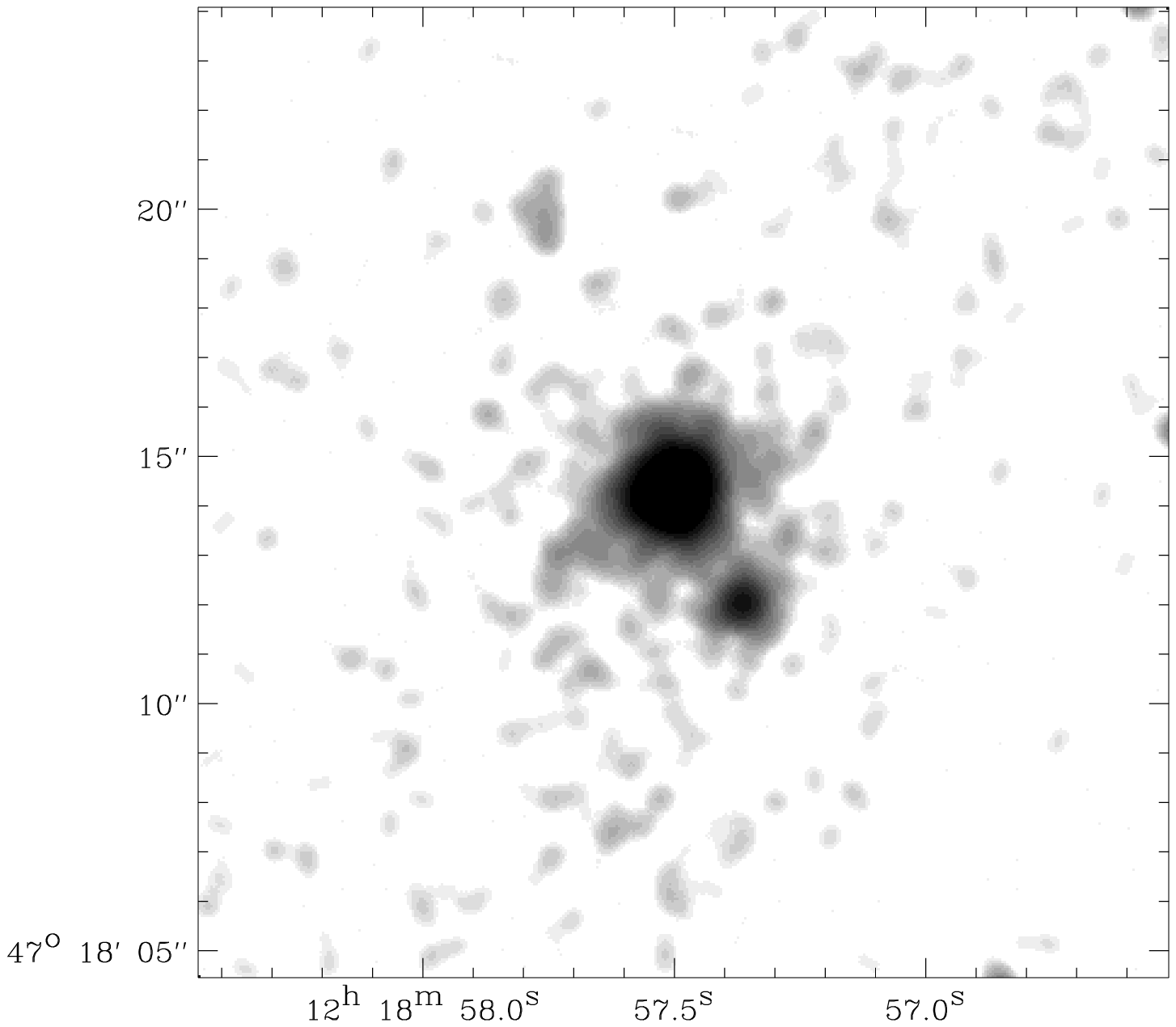}
  \hspace{-1.65cm}
  \includegraphics[scale=0.45]{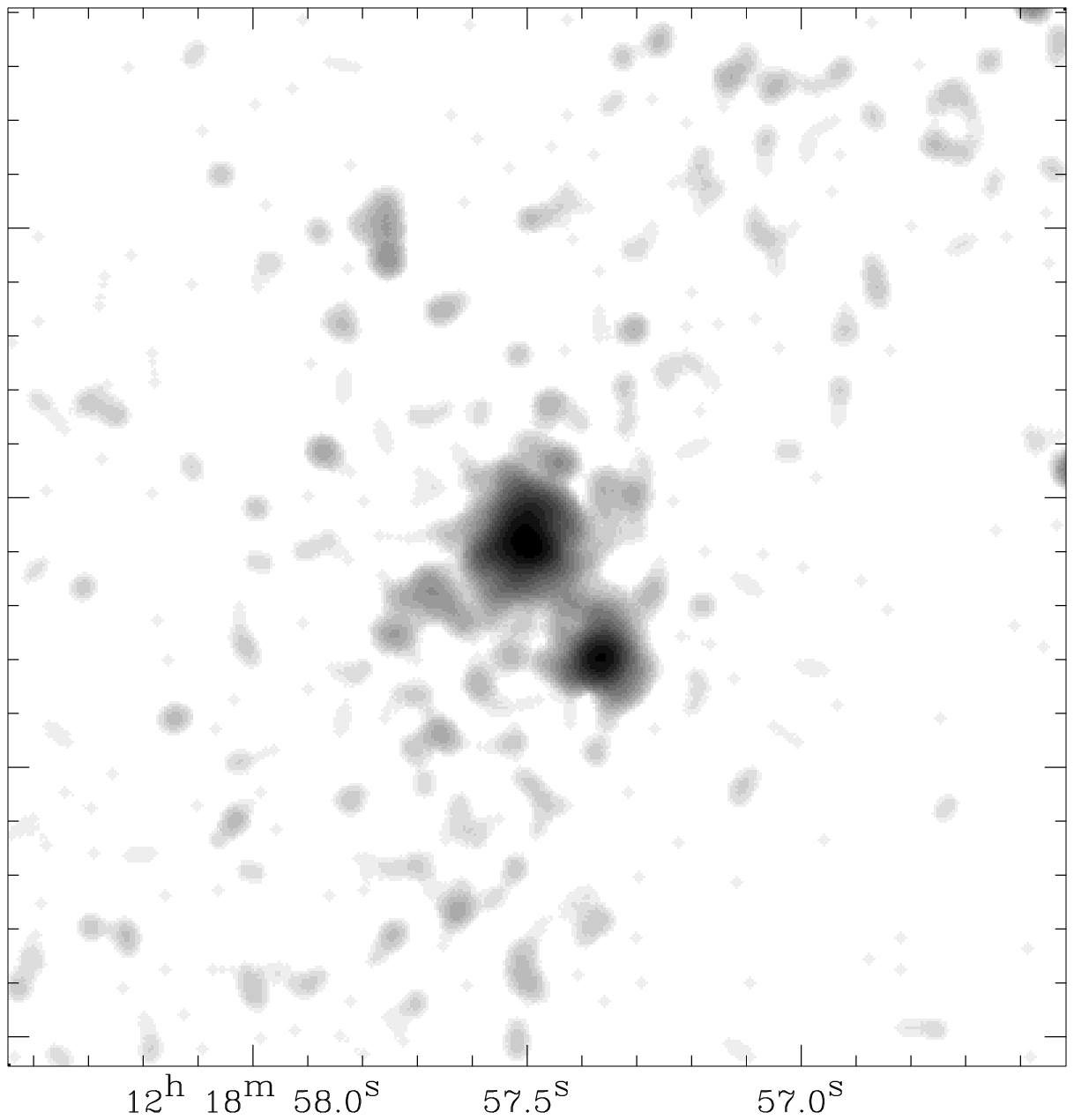}
  \hspace{-1.65cm}
  \includegraphics[scale=0.45]{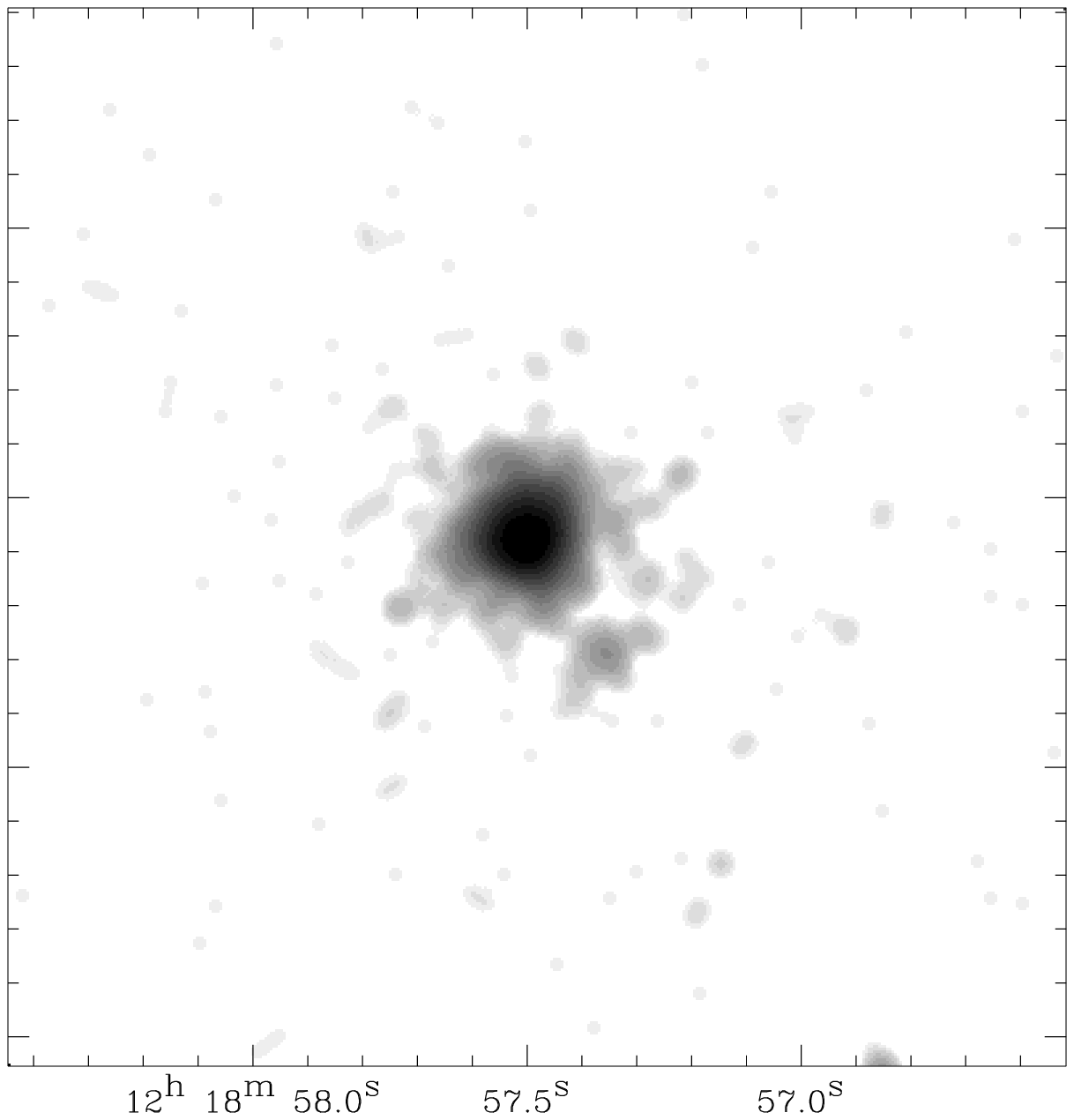}
}

\caption{Grey scale representations of Chandra X-ray images of the nucleus of
  NGC 4258 in the energy bands 0.5~--~8~keV (left), 0.5~--~3~keV (center) and
  3~--~8~keV (right). The images show the combined 3.2~sec frame-time data from
  the 13~ksec Cycle 1 and 20~ksec Cycle 2 observations. The images have been
  re-sampled to ten times smaller pixel size and smoothed by a Gaussian of FWHM
  $0\farcs5$. \label{fig:morph}}

\end{figure}

The 3.2 s frame-time observations of 2000-04-17 and 2001-05-28 have been
combined to produce the grey scale image shown in Fig.~\ref{fig:morph}. Here we
concentrate on the nucleus and its immediate environment, and do not discuss
the much larger scale ``anomalous arms'' (see WYC). As noted by WYC and by
\citet{2002A&A...384..793P}, there is a weaker source $2\farcs5$ southwest of
the nucleus. Soft, diffuse emission is seen to extend to the northwest and
southeast of the nucleus, and this emission is the beginning of the ``anomalous
arms''. Comparing the spatial distribution of the 3~--~8~keV counts from the
nucleus in the 2001-05-29 observation, which does not suffer from pile-up, with
a theoretical point spread function at 6.4~keV, we find that the hard X-ray
nucleus is unresolved.

\section{X-ray Light Curves} \label{sec:lightcurve}

\begin{figure}
\centerline{
  \includegraphics[scale=0.7,angle=270]{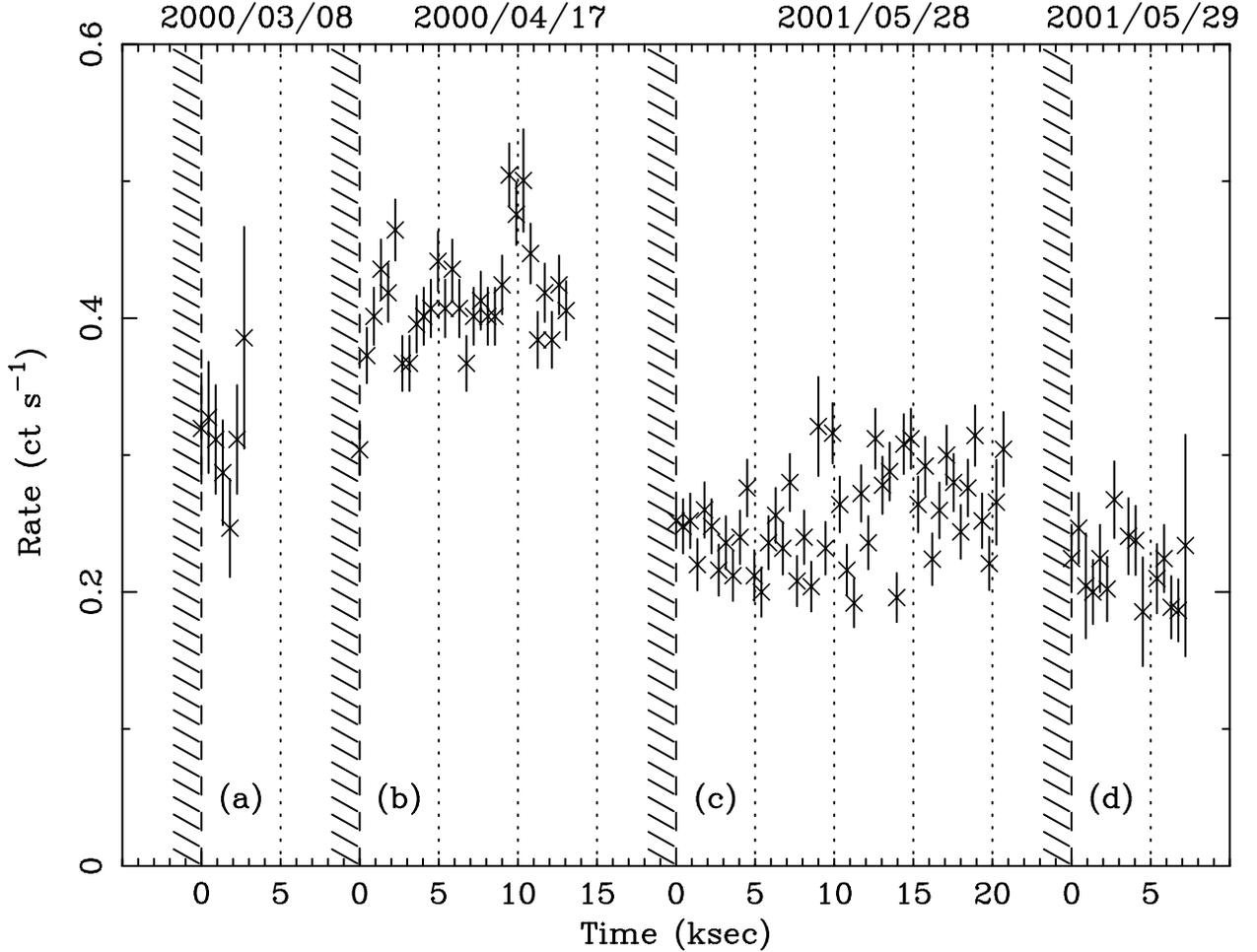}
}

\caption{X-ray light curves from circular apertures of radius $1\farcs5$
  centered on the nucleus of NGC 4258 in 450~s bins, from (a) the 1.6~ks
  0.4~sec frame-time observation (left), (b) the 13~ks 3.2~sec frame-time
  observation (center left), (c) the 20~ksec 3.2~sec frame-time observation
  (center right), and (d) the 7~ksec 0.4~sec frame-time observation (right).
  Vertical dotted lines are placed every 5~ksec, and the shaded regions
  indicate the start of a new observation. The 3.2~sec frame-time data sets,
  (b) and (c), have been corrected for the effects of pile-up. The background
  count rate is negligible ($< {\rm few} \times 10^{-4}$ cts s$^{-1}$). The
  error bars show the statistical noise, but do not include the systematic
  error associated with the pile-up correction of data sets (b) and (c).
  \label{fig:lc}}

\end{figure}

X-ray light curves of the nucleus of NGC 4258 in 450~s bins are shown in
Fig.~\ref{fig:lc}. The background count rate is negligible ($< {\rm few} \times
10^{-4}$ cts s$^{-1}$) and has not been subtracted. The 3.2~sec frame-time data
sets have been corrected for the effects of pile-up, using the continuum models
given in Section~\ref{sec:spec_nuc}. The light curves show 10~--~14\%\ rms
variability on short timescales and $\approxgt $20\%\ variability on year-long
timescales between the first and last observations (panels a and d, neither of
which is affected by pile-up). One of the effects of moderate to severe pile-up
is to reduce the sensitivity of the observed count rate to changes in the
actual count rate (i.e., as the actual count rate increases the observed count
rate increases at a much slower rate). This effect adds a systematic error to
our determinations of the un-piled mean count rates (panels b and c) as well as
reducing the observed amplitude of short timescale variability. This systematic
error is difficult to estimate, but may be as large as a ${\rm few} \times
10$\%.

\section{X-ray Spectra}

\subsection{Nucleus} \label{sec:spec_nuc}

\subsubsection{Continuum Spectrum} \label{sec:spec_nuc:cont}

\begin{figure}
\centerline{
  \includegraphics[scale=0.75,angle=270]{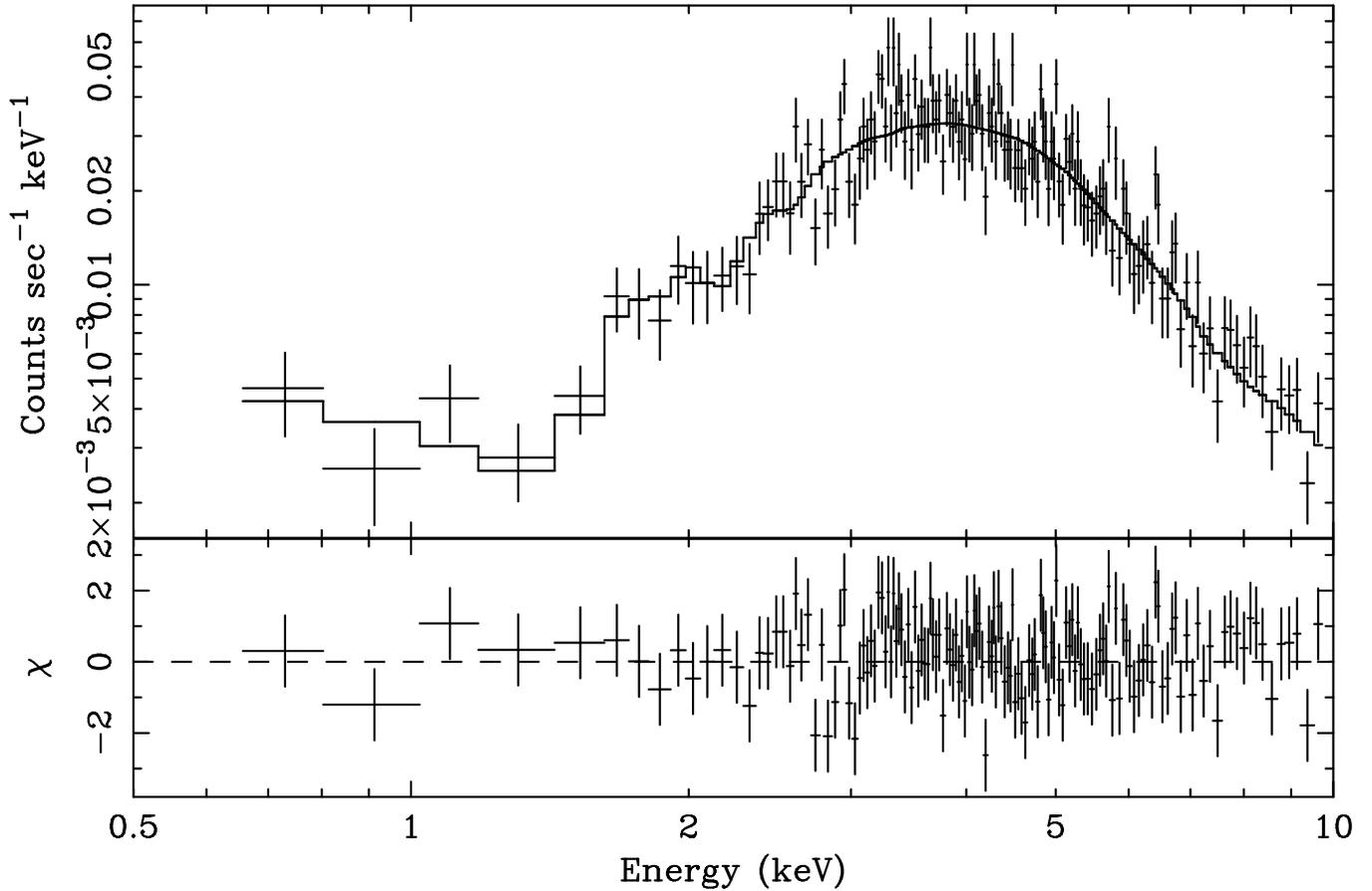}
}

\caption{X-ray spectrum of the nucleus of NGC 4258 extracted from the 20 ksec
  observation of 2001-05-28. The upper panel shows data points with error bars
  (crosses), with the model folded through the instrument response, taking into
  account the effects of pile-up (solid line). The lower panel shows the $\chi$
  residuals to this fit. The parameters of this model are listed in
  Table~\ref{tab:spec:continuum}. \label{fig:full_spec}}

\end{figure}

Initially we studied the 2001-05-29 observation (which has an exposure time of
7~ksec and a frame-time of 0.4~sec) because it is the highest S/N observation
not affected by pile-up. Two components are required to describe the spectrum
of the nucleus: i) a low energy component below $\sim 2$~keV, such as a 1~keV
bremsstrahlung absorbed by the Galactic column, and ii) a heavily absorbed
($N_{\rm H} \simeq 7 \times 10^{22}$ cm$^{-2}$) high energy component above
$\sim 2$~keV, such as a hard power law of photon index $\Gamma \simeq 1.4$ or a
very hot bremsstrahlung with a temperature $\approxgt 20$~keV. We then modeled
the other data sets in a similar manner, correcting for pile-up as necessary.
Table~\ref{tab:spec:continuum} summarizes our models of the continuous spectra
of the nucleus for the different observations, and Fig.~\ref{fig:full_spec}
shows a model fit to the 2001-05-28 data set. For each of these observations,
the soft X-ray component is the same while the hard X-ray component declined
from 2000 to 2001. During this variation, the intrinsic absorbing column
density and photon index of the hard component remained constant
(Table~\ref{tab:spec:continuum}).  The models presented here assume absorption
by a single column of neutral gas.  If the absorber is ionized then a higher
column density may be permitted (see discussion in Sections \ref{sec:photo} and
\ref{sec:collision}).  If a partial covering model is used the best fit has a
covering fraction of $\approxgt$ 99\%\ and parameters very similar to our
single absorber fits;  the uncovered fraction, $\approxlt$ 1\%\ of the nucleus,
accounts for the observed soft emission.  It is also possible for an additional
absorption component to be added to our models;  $\sim$ 40\%\ of the nucleus
may be absorbed by $N_{\rm H} \sim 7 \times 10^{22}$ cm$^{-2}$ while $\sim$
60\%\ of the nucleus may be absorbed by an additional $N_{\rm H} \sim 8 \times
10^{23}$ cm$^{-2}$, and this model requires a softer power law, with $\Gamma
\sim 1.9$.  In addition, an arbitrary amount of complete obscuration may be
present that we cannot constrain.  Our decision to use a single column density
is the most conservative approach and this column density is a lower limit to
the true column density.

\subsubsection{Iron Emission Lines}

\begin{figure}
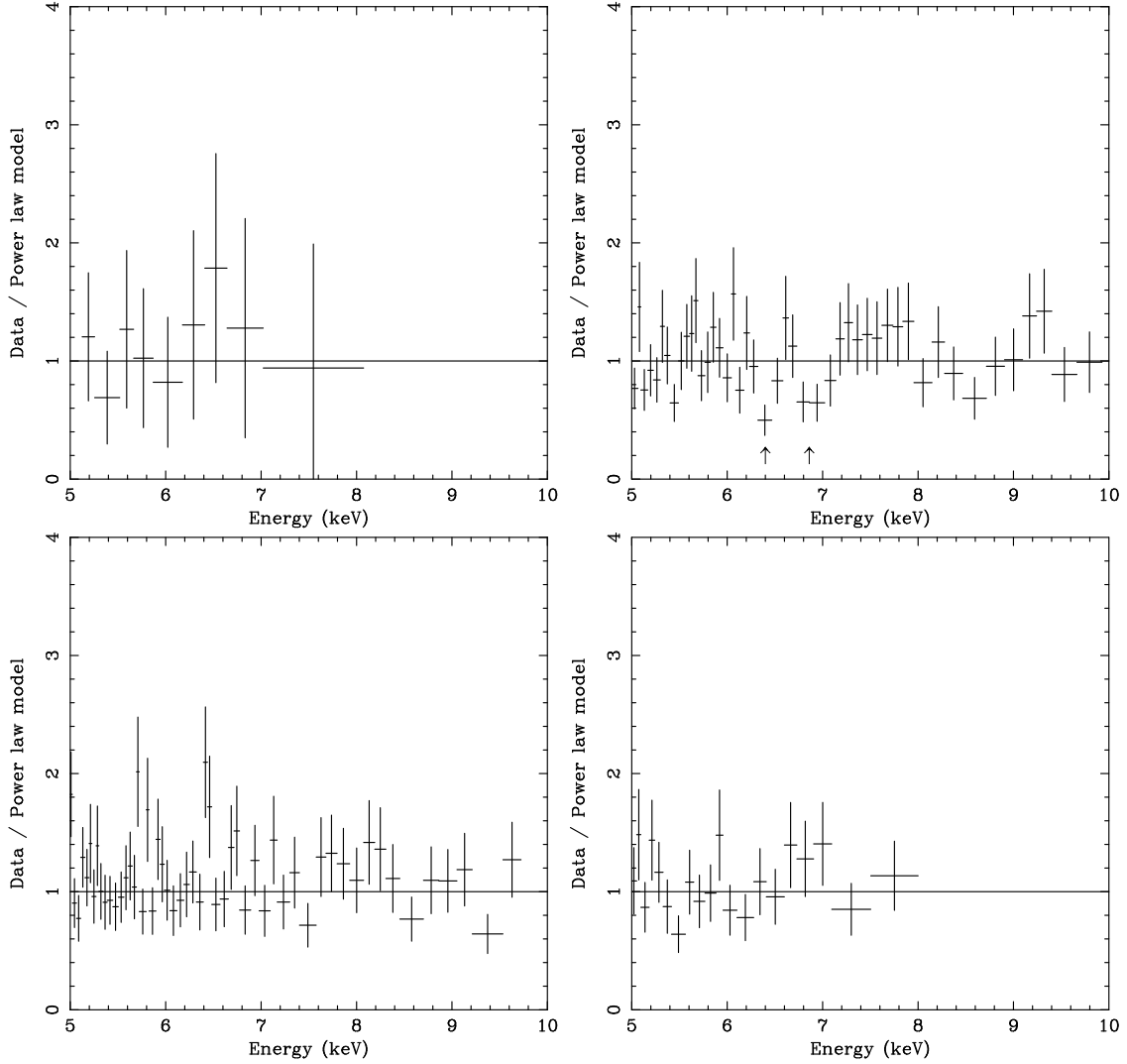


\centerline{
  \includegraphics[scale=0.4,angle=270]{fig4a.ps}
  \includegraphics[scale=0.4,angle=270]{fig4b.ps}
}
\centerline{
  \includegraphics[scale=0.4,angle=270]{fig4c.ps}
  \includegraphics[scale=0.4,angle=270]{fig4d.ps}
}

\caption{Chandra spectra of the nucleus of NGC 4258 including the Fe K$\alpha$
  line region.  Observations were made on 2000-03-08 for 1.6~ksec with a
  0.4~sec frame-time (upper left panel), 2000-04-17 for 13~ksec with a 3.2~sec
  frame-time (upper right panel), 2001-05-28 for 20~ksec with a 3.2~sec
  frame-time (lower left panel) and 2001-05-29 for 7~ksec with a 0.4~sec
  frame-time (lower right panel).  The spectra are grouped to have $\ge 15$ cts
  bin$^{-1}$ and the energy scale is as observed.  The plots show the ratio of
  the data to the best fitting absorbed power law model above 5~keV.  There is
  no evidence of strong Fe K$\alpha$ emission.  On 2000-04-17, statistically
  significant absorption features are seen at 6.4~keV and 6.9~keV and are
  indicated by the arrows in the upper right panel.  \label{fig:fe}}

\end{figure}

We now focus our attention on the spectrum in the region of the Fe K$\alpha$
lines.  The ratios of the data to the best fitting power law models above 5~keV
are shown in Fig.~\ref{fig:fe}, in which no strong iron K$\alpha$ emission
lines are apparent.  If a narrow, neutral iron K$\alpha$ emission line at a
rest frame energy of 6.4~keV is added to the model, the upper limit to the EW
of such a line is ranges between 94 and 887~eV (90\%\ confidence) for the
different observations (see  Table~\ref{tab:fe_lines}).  If the data are
grouped to have $\ge 10$ or $\ge 3$ counts per bin, there is tentative evidence
of iron K$\alpha$ emission lines in the first observation and the last two
observations at the energies expected for neutral and H-like iron, although
these features have low statistical significance.

\subsubsection{Iron Absorption Lines} \label{sec:fe_abs_lines}

During the 2000-04-17 observation two strong absorption features are seen at
NGC 4258 rest frame energies of $6.41^{+0.05}_{-0.06}$~keV and
$6.87^{+0.09}_{-0.06}$~keV, with equivalent widths of $-166$~eV and $-215$~eV,
respectively (see Fig.~\ref{fig:fe} and Table~\ref{tab:fe_lines}; the
heliocentric redshift of NGC 4258 is $z = 0.00155$
\citep{1992ApJ...390..365C}).  These absorption lines at 6.41~keV and 6.87~keV
are both statistically significant at $>$ 99.5\%\ confidence
(Table~\ref{tab:fe_lines}).  An F-test may be used to quantify the improvement
to the simple power law model when the absorption features are added.  The
power law model has $\chi^2$ / degrees of freedom (dof) of 115.8 / 101.  Adding
a single absorption line at 6.41 keV (the energies of all absorption features
discussed here have been left as free parameters) to the power law model gives
an improvement of $\Delta \chi^2$ / $\Delta$ dof = 12.9 / 2, which is
significant with 99.5\%\ confidence.  Adding a single absorption line at 6.87
keV to the power law model gives an improvement of $\Delta \chi^2$ / $\Delta$
dof = 7.7 / 2, which is significant with 96.2\%\ confidence, although the
absence of the 6.41 keV line is so significant that it affects the continuum
fit in the iron line band.  If we add the 6.87 keV line to the power law plus
6.41 keV line model we find an improvement with the addition of the 6.9 keV
line of $\Delta \chi^2$ / $\Delta$ dof = 11.6 / 2, which is significant with
99.5\%\ confidence.  If we compare the power law model to the model with two
absorption lines we find $\Delta \chi^2$ / $\Delta$ dof = 23.5 / 4, significant
at $>$ 99.9\%.  Only three of our observations have sufficient signal-to-noise
ratio to detect absorption lines of this strength, and if we assume they are
either ``cold'' lines at $\sim 6.4$ keV or ``hot'' lines at $\sim 6.9$ keV,
this give us only six ``chances'' to see a strong iron absorption feature, so
the presence of absorption lines at these energies in this data set is
significant with at least 99.5\%\ confidence.  Continuing to assume the
absorption lines are subject to only the redshift of NGC 4258, the 6.41~keV
line is consistent with either of or a blend of the $n = 1 \rightarrow 2$
resonance lines of Fe XVIII and Fe XIX at 6.43 and 6.46 keV, respectively (the
energy expected for Fe XX is 6.50 keV and inconsistent with the data, while
ionization states below Fe XVIII do not have an L-shell vacancy).  The feature
at 6.87~keV is probably the $n = 1 \rightarrow 2$ resonance line of Fe XXVI at
6.97 keV (the energy expected of Fe XXV is 6.70 keV which is inconsistent with
the measured value).

There is no evidence of a strong iron K absorption edge at 7.8 keV \citep[from
Fe XVIII; we use the edge energies given by][]{1995A&AS..109..125V}, 7.9 keV
(from Fe XIX), with a limit of $\tau_{\rm edge} \approxlt 0.5$.  We are
insensitive to an absorption edge at 9.3 keV (from Fe XXVI).

\begin{figure}
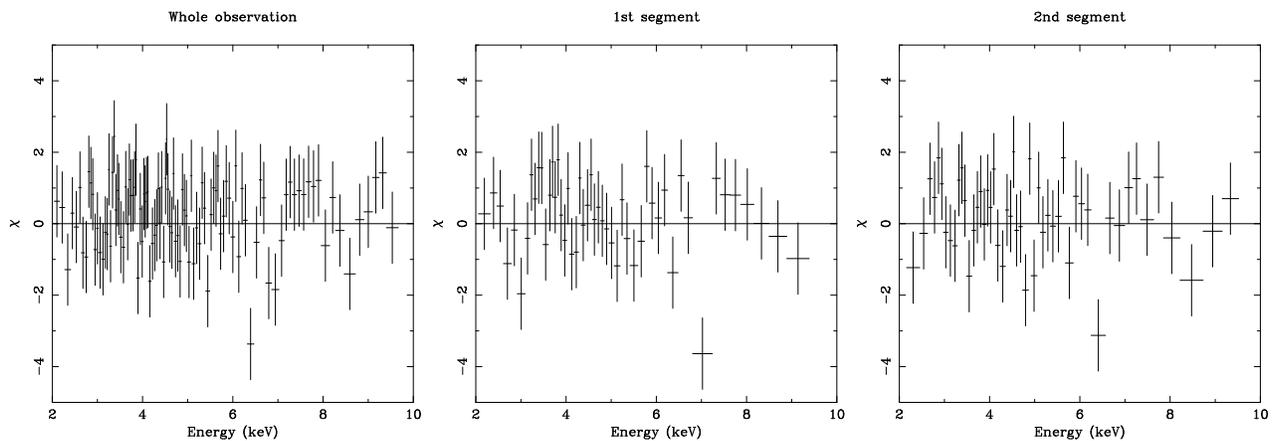

\centerline{
  \includegraphics[scale=0.3,angle=270]{fig5a.ps}
  \includegraphics[scale=0.3,angle=270]{fig5b.ps}
  \includegraphics[scale=0.3,angle=270]{fig5c.ps}
}

\caption{Chandra spectra of the nucleus of NGC 4258 covering the Fe K$\alpha$
  line region from the 2000-04-17 13 ksec observation.  The figures show the
  residuals, in terms of $\chi$, from a power law model for the entire
  observation (left panel), the first 6000 sec segment of the observation
  (center panel) and the second 6000 sec segment of the observation (right
  panel).  The presence of a strong absorption feature at 6.9 keV in the first
  segment is highly significant (at $>$ 99.5\%\ confidence), as is its
  disappearance in the subsequent segment.  The 6.4~keV absorption line is
  marginally stronger in the second segment, where its presence is significant
  at $>$ 99.5\%\ confidence.  The absorption line parameters for each panel are
  given in Table~\ref{tab:fe_lines}.  \label{fig:fe_line_segments}}

\end{figure}

\label{sec:unusual} If the 2000-04-17 observation is divided into two
consecutive segments, each 6 ksec in duration, we can attempt to investigate
any short timescale variability of the absorption features.  The spectra of the
iron line region for each of these segments is shown in
Fig.~\ref{fig:fe_line_segments}. A strong absorption line is seen at $\sim
6.9$~keV in the first segment, but has disappeared in the second segment
6000~sec later. In the first segment, the addition of a narrow absorption line
at $6.95^{+0.22}_{-0.12}$~keV to the continuum model is statistically
significant with $>$ 99.5\%\ confidence (see Table~\ref{tab:fe_lines}).  In the
second segment, the addition of an absorption feature at 6.9 keV to the
continuum model does not improve the fit at all.  If the model of the second
segment includes an absorption line at 6.95~keV, identical to that observed in
the first segment, we find that removing this line improves the quality of the
fit with $>$ 95\%\ confidence.  These considerations indicate that the narrow
absorption feature at 6.9 keV in the first segment is real and that it is
significantly weaker in the second segment.  The absorption feature at 6.4 keV
is seen during both segments; in the first, it is significant at only the $>$
90\%\ confidence level, but is marginally stronger in the second segment where
it is statistically significant at $>$ 99.5\%\ confidence.

\subsection{Off-nuclear Source}

The spectrum of the off-nuclear source taken from the 20~ksec observation on
2001-05-28 is well described by an absorbed power law. The intrinsic column
density is $N_{\rm H} = 4.3^{+2.8}_{-1.6} \times 10^{21}$ cm$^{-2}$, the power
law photon index is $\Gamma = 1.8^{+0.5}_{-0.4}$, and the unabsorbed
2~--~10~keV luminosity is $L_{\rm x} = 6.1 \times 10^{38}$ erg s$^{-1}$
(assuming the source is in NGC 4258). These parameters are consistent with
those found by WYC from the observation on 2000-04-17.

\section{The Fe Absorption Lines}

\subsection{Curve of Growth Analysis}

\begin{figure}
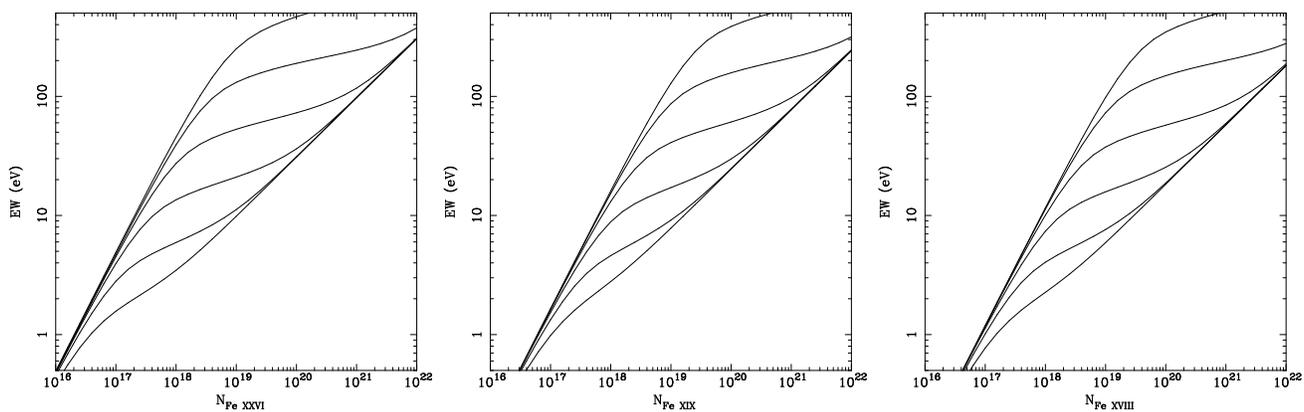

\centerline{
  \includegraphics[scale=0.3,angle=270]{fig6a.ps}
  \includegraphics[scale=0.3,angle=270]{fig6b.ps}
  \includegraphics[scale=0.3,angle=270]{fig6c.ps}
}

\caption{Curves of growth for Fe XXVI K$\alpha$ (left panel), Fe XIX K$\alpha$
  (center panel) and Fe XVIII K$\alpha$ (right panel).  The curves in each
  panel correspond to kinematic temperatures of 10,000 keV, 1000 keV, 100 keV,
  10 keV, 1 keV and 0.1 keV, from top to bottom, respectively.  The $1 \sigma$
  line of sight velocity dispersion of the iron ions is $\Delta v \sim 4.1
  \times 10^8$ cm s$^{-1}$ at 10,000 keV and $\Delta v \sim 1.3 \times 10^6$ cm
  s$^{-1}$ at 0.1 keV. \label{fig:cog}}

\end{figure}

In order to see a strong absorption line, the solid angle of the resonantly
scattering material as seen by the central X-ray source must be small,
otherwise the absorption line would be ``filled in'' by emission lines. We thus
imagine the absorbing material as a ``blob'' of gas, and use the strength of an
absorption line to estimate the column density of the iron ion responsible for
the absorption. Following the method described by \citet{2000ApJ...539..413K}
\citep[also see][]{1978ppim.book.....S}, the equivalent width (EW), $W_\nu$, of
a resonance line at a frequency $\nu$ is given by \begin{equation} W_\nu =
\int^{\infty}_{0} \left\{ 1 - e^{-N_{\rm Fe}\sigma(\nu)} \right\} d\nu {\rm
~}[{\rm Hz}] \end{equation} where $N_{\rm Fe}$ is the column density of the
iron ion in question and $\sigma(\nu)$ is the cross section of that ion at a
frequency $\nu$. If the ions obey a Maxwell-Boltzmann distribution, then
$\sigma(\nu)$ has a Voigt profile that depends on the temperature of the blob,
being broader at a higher temperature. If the energy distribution of the ions
is dominated by turbulence or bulk motion, then the appropriate turbulent or
bulk velocity dispersion is used in place of the temperature-dependent velocity
dispersion.  The oscillator strengths and Einstein coefficients required to
determine the cross sections were taken from \citet{1996ADNDT..64....1V} for Fe
XXVI K$\alpha$ and \citet{2002ApJ...570..165B} for Fe XVIII and Fe XIX
K$\alpha$. Equation (1) was integrated for different values of $N_{\rm Fe}$
giving the ``curve of growth'' ($W_\nu$ vs. $N_{\rm Fe}$), shown in
Fig.~\ref{fig:cog} for Fe XXVI \citep[cf.][their Fig. 3]{2000ApJ...539..413K},
Fe XVIII and Fe XIX K$\alpha$ (the absorption line at $\sim 6.4$ keV may be a
blend of Fe XVIII and Fe XIX, and here we assume that the iron ions are equally
divided between Fe XVIII and Fe XIX). From Fig.~\ref{fig:cog}, we note that the
large EW of the absorption lines we see ($-166$ -- $-215$ eV;
Table~\ref{tab:fe_lines}) can correspond to a broad range of iron column
densities depending on the kinematic temperature of the blob. For example, the
column density of Fe XXVI required to produce a 215 eV EW absorption line
decreases from $N_{\rm Fe~XXVI} = 6 \times 10^{21}$ cm$^{-2}$ at a temperature
of $kT \le 100$ keV (equivalent to a $1 \sigma$ line of sight velocity
dispersion of the iron ions of $\Delta v \sim 4.1 \times 10^7$ cm s$^{-1}$) to
$N_{\rm Fe~XXVI} = 5 \times 10^{18}$ cm$^{-2}$ at $kT = 10,000$ keV (equivalent
to a $1 \sigma$ velocity dispersion of the iron ions of $\Delta v \sim 4.1
\times 10^8$ cm s$^{-1}$).

A complication is that, while we see strong absorption lines from a blend of Fe
XVIII -- Fe XIX and from Fe XXVI, we do not see a strong absorption line from
Fe XXV. For a single temperature, collisionally ionized or photoionized plasma,
such is not possible \citep[see][respectively]{1992ApJ...398..394A,
1984ApJ...286..366K}, so there must be two distinct blobs with different
ionization states. Since the absorption line from Fe XVIII -- Fe XIX did not
vary during the 13 ksec exposure while the Fe XXVI absorption line did vary, it
is likely that the $\sim 6.4$ keV absorption line originates in cooler gas at
larger radii. We shall, therefore, assume that the $\sim 6.4$ keV and $\sim
6.9$ keV absorption lines originate in spatially distinct blobs.  We also
assumed that the ion velocities within each absorbing blob are well described
by a Maxwell-Boltzmann distribution since the CCD-resolution spectra provide
little information about the line profile.  This need not be the case, and the
absorption lines may be made up of a number of individual components some of
which may be saturated while others are unsaturated (equivalently the blob may
be made up of sub-blobs of different column densities and velocity
dispersions).  It is difficult to quantify how these differences would affect
our analysis, but the velocity dispersion that we derive can be thought of as a
column density weighted mean velocity dispersion of (and within) the sub-blobs.

In the following, we derive some physical parameters of the ``blobs''. In
particular, we are interested in their opacity in the 2 -- 5 keV region, as
this is the range of the spectrum from which the equivalent hydrogen absorbing
column of $N_{\rm H} = 7 \times 10^{22} \pcmsq$ was derived, assuming the gas
is neutral and of solar abundance (Section \ref{sec:spec_nuc:cont}).

\subsubsection{Photoionized Plasma} \label{sec:photo}

We first investigate the blob responsible for the Fe XXVI K$\alpha$ absorption
line.  For $W_\nu = -215$ eV (Table \ref{tab:fe_lines}), our curve of growth
analysis indicates that $N_{\rm Fe~XXVI} \approxgt 6 \times 10^{18}$ cm$^{-2}$
(see Fig.~\ref{fig:cog}) and, if the iron abundance is equal to the solar value
of $3.5 \times 10^{-5}$ Fe ions per hydrogen nucleus
\citep{1989GeCoA..53..197A}, this column corresponds to $N_{\rm H} ({\rm
ionized}) \approxgt 2 \times 10^{23}$ cm$^{-2}$.  If the gas in this blob is
photoionized so that the ionization state of iron is dominated by Fe XXVI, an
ionization parameter $\xi = L_{\rm x} / (n_e R^2) \simeq 3000$ erg cm s$^{-1}$
is required \citep{1982ApJS...50..263K}, where $n_e$ is the electron density
and $R$ is the distance of the blob from the ionizing source. The plasma
temperature corresponding to $\xi = 3000$ erg cm s$^{-1}$ is $T \simeq 3 \times
10^6$ K, and $\simeq$ 60\%\ of the iron ions are in the form of Fe XXVI, with
the remainder being evenly split between Fe XXV and Fe XXVII
\citep{1982ApJS...50..263K}. The X-ray opacity of such a gas is reduced by a
large factor below the Fe K edge \citep{1984ApJ...286..366K} and hence a column
density of $N_{\rm H} ({\rm ionized}) \approxgt 2 \times 10^{23}$ cm$^{-2}$ is
consistent with $N_{\rm H} ({\rm obs}) = 7 \times 10^{22}$ cm$^{-2}$.  If we
consider instead the stronger 6.9~keV line seen during the first segment of the
2000-04-17 observation the column density requirements are increased by a
factor of a few, and are still consistent with $N_{\rm H} ({\rm obs})$.

We now turn our attention to the blob producing the Fe K$\alpha$ 6.4 keV
absorption feature from a blend of Fe XVIII--Fe XIX K$\alpha$.  For $W_\nu =
-166$~eV (Table~\ref{tab:fe_lines}) our curve of growth analysis indicates that
$N_{\rm Fe} \approxgt 10^{19}$ cm$^{-2}$ (see Fig.~\ref{fig:cog}) and, if the
iron abundance is equal to the solar value, this corresponds to $N_{\rm H}
({\rm ionized}) \approxgt 3 \times 10^{23}$ cm$^{-2}$.  A photoionized plasma
with an ionization parameter of $\xi \simeq 100$ erg cm s$^{-1}$ has almost all
of its iron ions evenly distributed between Fe XVIII and Fe XIX \citep[][ model
1]{1982ApJS...50..263K}. The plasma temperature corresponding to $\xi = 100$
erg cm s$^{-1}$ is $T \simeq 10^5$ K. From 2 to 5 keV the opacity is $\simeq 3$
times smaller than that of a neutral gas \citep{1984ApJ...286..366K}.  Given
the approximations inherent to our analysis, the required column density of
$N_{\rm H} ({\rm ionized}) \approxgt 3 \times 10^{23}$ cm$^{-2}$ is consistent
with the equivalent hydrogen column density for neutral gas ($N_{\rm H} = 7
\times 10^{22}$ cm$^{-2}$ -- see Table~\ref{tab:spec:continuum}) needed to
account for the observed \emph{photoelectric} absorption.  This consistency,
however, requires a very high kinematic temperature of $kT \approxgt 1000$ keV,
which presumably represents bulk motion with a $1 \sigma$ line of sight
velocity dispersion of the iron ions of $\Delta v \approxgt 1.3 \times 10^8$ cm
s$^{-1}$.  We conclude that the gas responsible for the observed photoelectric
absorption may be the same as that responsible for the 6.4~keV Fe absorption
line.

\subsubsection{Collisionally ionized Plasma} \label{sec:collision}

We first consider the blob responsible for the Fe XXVI K$\alpha$ absorption
line. If the absorbing gas is collisionally ionized, the ionic fraction of Fe
XXVI is maximal at a temperature $T \simeq 10^8 \K$
\citep{1992ApJ...398..394A}, and such a plasma will have negligible opacity in
the 2 -- 5 keV band \citep{1984ApJ...286..366K}.

Considering next the blob responsible for the $\sim 6.4$ keV absorption line,
the ionic species Fe XVIII -- Fe XIX have maximal fractions at a temperature of
$T \simeq 10^7$ K \citep{1992ApJ...398..394A} and such a plasma will have a
negligible opacity at energies $\approxlt 2$ keV, and roughly an order of
magnitude lower opacity than neutral gas in the 2 -- 5 keV band
\citep{1984ApJ...286..366K}. If such a blob is responsible for \emph{both} the
observed photoelectric absorption and the $-166$~eV EW absorption line at
6.4~keV, a kinematic temperature of $kT \sim 1000 \keV$ (equivalent to a $1
\sigma$ line of sight velocity dispersion of the iron ions of $\Delta v \sim
1.3 \times 10^8 \cmps$) is needed.  In this case, the absorbing gas could be
in collisional equilibrium at $10^7 \K$, with the above value of $\Delta v$
representing the turbulent or bulk outflow velocity.  Thus, again, it is
possible that the gas responsible for the observed photoelectric absorption is
the same as that which gives the Fe K$\alpha$ absorption line at 6.4 keV.

We emphasize that all other observations of NGC 4258 made by us
(Table~\ref{tab:fe_lines}) and others (Section~\ref{sec:intro}) have found
either no Fe lines or Fe lines in emission.  Thus the spectrum is clearly
variable \citep[cf.][]{2002ApJS..139....1T}.  If we adopt the conclusion that
the gas responsible for the Fe K$\alpha$ 6.4 keV absorption seen on 2000-04-17
is indeed the same as that which causes the photoelectric absorption, then the
variability at 6.4~keV would result from variability in the emission line,
since the photoelectric absorbing column density appears to be constant
(Table~\ref{tab:spec:continuum} and papers cited in Section~\ref{sec:intro}).

\subsubsection{Nature of Absorbing Plasma}

Since our line of sight to the nucleus grazes the surface of the thin, masing
disk of NGC 4258 \citep[inclination $i = 82\degmark \pm 1\degmark$;
][]{1999Natur.400..539H}, the absorbing blobs may be part of a disk-wind
boundary layer or an outflow.  The absence of strong iron K$\alpha$ emission is
consistent with a hot, low radiative efficiency, advection dominated accretion
flow \citep[ADAF; e.g.,][]{1994ApJ...428L..13N, 1995ApJ...438L..37A} with a
lower limit to the radius of an outer thin disk of approximately $100 R_G$
\citep{2000ApJ...540..143R}.  Such accretion flows may show strong outflows
\citep{1999MNRAS.303L...1B} or convection \citep{2000ApJ...539..809Q,
2000ApJ...539..798N} and the ionized absorbing ``blobs'' we have found in NGC
4258 may be inhomogeneities in such a flow with high turbulent velocities (to
account for the large equivalent widths of the absorption lines -- see
Fig.~\ref{fig:cog}).  The short-timescale variability of the 6.9~keV absorption
line (Section~\ref{sec:fe_abs_lines}) might result from absorption by a blob at
$\sim 10 R_G$ which corresponds to a dynamical timescale of $\sim 6000$~sec.

The X-ray emission from the nucleus of NGC 4258 may originate from the base of
the jet. As gas transitions from the ADAF to the jet it is shock heated, and
the X-ray continuum emission may be synchrotron self-Compton emission from the
accelerated electrons \citep{2002A&A...391..139Y}. The mass ejection process
may be unstable and the absorption lines we see could be from blobs of cooling
thermal gas entrained in the jet outflow.

It is also very interesting to note that similar X-ray absorption features have
been seen from the Galactic black hole or neutron star binaries GRO J1655-40
\citep{1998ApJ...492..782U, 2001PASJ...53..179Y}, Circinus X--1
\citep{2000ApJ...544L.123B, 2002ApJ...572..971S}, GRS 1915+105
\citep{2000ApJ...539..413K, 2002ApJ...567.1102L}, GX 13+1
\citep{2001ApJ...556L..87U}, MXB 1659-298 \citep{2001A&A...379..540S}, X
1624-490 \citep{2002A&A...386..910P} and X 1254-690
\citep{2003astro.ph..6526B}, all of which have high inclination accretion disks
($i \simeq 60\degmark$ -- $80\degmark$, with the exception of GX 13+1 for which
the inclination is not known).  If a high inclination is required to produce a
strong absorption line, as is suggested by these observations, then NGC 4258 is
an important galaxy to study the connection between accretion disks in AGN and
Galactic systems.

\section{Conclusions}

Our four Chandra observations of the low luminosity active galactic nucleus of
NGC 4258 have shown:

i) The continuum spectrum contains a component below 2 keV, such as 1 keV
thermal emission absorbed by the Galactic column, plus a component above 2 keV,
which is well described by a hard power law ($\Gamma = 1.4$) absorbed by an
equivalent hydrogen column of $N_{\rm H} = 7 \times 10^{22}$ cm$^{-2}$.

ii) The soft component of the continuum spectrum does not vary, while the flux
of the hard component declined from March/April 2000 to May 2001. During this
decline, the intrinsic absorbing column density and photon index of the hard
component remained constant (Table~\ref{tab:spec:continuum}).

iii)  We do not detect an iron K$\alpha$ emission line, with the upper limit to
the EW of a narrow, neutral iron line ranging between 94 and 887 eV (90\%\
confidence) for the different observations (Table~\ref{tab:fe_lines}).  In the
second observation (on 2000-04-17) absorption lines are seen at 6.4 and 6.9 keV
that are statistically significant with $>$ 99.5\%\ confidence. There is
evidence that the absorption line at 6.9 keV varied on a timescale of 6000
secs.

iv) We argue that the absorption lines are Fe K$\alpha$ resonant absorption,
with that at 6.4 keV by Fe XVIII -- Fe XIX, and that at 6.9 keV by Fe XXVI
ions. A curve of growth analysis has been performed for each line in order to
obtain the relationship between the kinematic temperature and the ionic column
for the observed equivalent width.

v) Given that the maser disk is viewed at an inclination $i = 82\degmark$, we
suggest that the 6.9 keV absorbing gas may be in an inner disk, a disk-wind
boundary layer or be thermal gas entrained at the base of the jet.

vi) The gas responsible for the 6.4 keV absorption line may be photoionized or
collisionally ionized. In each case, either the ionization temperature is lower
than the kinematic temperature  or (and more likely) the absorbing matter has a
high bulk velocity dispersion ($1 \sigma$ line of sight velocity dispersion
$\Delta v \simeq 1.3 \times 10^8$ cm s$^{-1}$). We have shown that this gas may
be the same as that which is responsible for the observed photoelectric
absorption.

Further X-ray observations of the nucleus of NGC 4258 are required to monitor
the variability of the iron absorption and emission lines. Such observations
should have higher sensitivity and better spectral resolution than those
presented here, with the goal of clean separation of absorption and emission
lines.

\acknowledgments

We thank Yuichi Terashima and Chris Reynolds for discussions. This research was
supported by NASA through grants NAG 8-1027 and NAG 8-1755.

\bibliographystyle{apj}
\bibliography{apj-jour,bibliography}

%%%%%%%%%%%%%%%%%%%%%%%%%%%%%%%%%%%%%%%%%%%%%%%%%%%%%%%%%%%%%%%%%%%%%%

\vfil\eject\clearpage
\begin{deluxetable}{cccccccc}
\tablewidth{0pt} %\rotate
  
\tablecaption{Chandra Observations of NGC 4258}

\tablecolumns{7}

\tablehead{
  \colhead{Date of obs.} &
  \colhead{Frame-time} &
  \colhead{Exposure time\tablenotemark{a}} &
  \colhead{Observed count rate} &
  \colhead{De-piled count rate} \\
  
  \colhead{} &
  \colhead{(sec)} &
  \colhead{(sec)} &
  \colhead{(counts sec$^{-1}$)} &
  \colhead{(counts sec$^{-1}$)}
}

\startdata

2000-03-08 & 0.4 & 1570 & $0.304 \pm 0.014$ & $0.304 \pm 0.014$ \\

2000-04-17 & 3.2 & 13056 & $0.161 \pm 0.004$ & 0.43\tablenotemark{b} \\

2001-05-28 & 3.2 & 20214 & $0.143 \pm 0.003$ & 0.28\tablenotemark{b} \\

2001-05-29 & 0.4 & 6819 & $0.243 \pm 0.006$ & $0.243 \pm 0.006$ \\

\enddata

\tablenotetext{a}{Total good exposure time, taking into account the good time
  intervals and dead-time correction factor.}

\tablenotetext{b}{These count rates are subject to increased errors (see
  section~\ref{sec:lightcurve}).}

\label{tab:data}
\end{deluxetable}

%%%%%%%%%%%%%%%%%%%%%%%%%%%%%%%%%%%%%%%%%%%%%%%%%%%%%%%%%%%%%%%%%%%%%%

\vfil\eject\clearpage
\begin{deluxetable}{ccccccccc}
\tablewidth{0pt} \rotate
  
\tablecaption{Continuum Spectral Models\tablenotemark{a} ~of the Nucleus of
  NGC 4258.}

\tablecolumns{8}

\tablehead{
  \colhead{Date of obs.} &
  \colhead{$kT_l$} &
  \colhead{$L_l$\tablenotemark{b}} &
  \colhead{$N_{\rm H}$\tablenotemark{c}} &
  \colhead{$\Gamma_{\rm h}$} &
  \colhead{$kT_h$} &
  \colhead{$L_h$\tablenotemark{d}} &
  \colhead{$\chi^2$ / dof} \\
  
  \colhead{} &
  \colhead{(keV)} &
  \colhead{(erg s$^{-1}$)} &
  \colhead{(cm$^{-2}$)} &
  \colhead{} &
  \colhead{(keV)} &
  \colhead{(erg s$^{-1}$)} &
  \colhead{}
}

\startdata

2000-03-08 & n/a & n/a & $6.9^{+2.2}_{-1.8} \times 10^{22}$ &
$1.3^{+0.6}_{-0.6}$ & & $1.2 \times 10^{41}$ & 27 / 26 \\

2000-04-17\tablenotemark{e} & $1.3^{+\infty}_{-0.9}$ & $1.6 \times 10^{38}$ &
$7.2^{+0.7}_{-0.4} \times 10^{22}$ & $1.5^{+0.1}_{-0.0}$ & & $1.3 \times
10^{41}$ & 113 / 103 \\

2001-05-28\tablenotemark{e} & $1.3^{+\infty}_{-0.8}$ & $1.4 \times 10^{38}$ &
$6.5^{+0.6}_{-0.3} \times 10^{22}$ & $1.5^{+0.1}_{-0.1}$ & & $7.9 \times
10^{40}$ & 158 / 143 \\

2001-05-29 & $1.0^{+0.6}_{-0.7}$ & $1.5 \times 10^{38}$ & $6.8^{+1.2}_{-1.0}
\times 10^{22}$ & $1.4^{+0.3}_{-0.3}$ & & $7.4 \times 10^{40}$ & 105 / 93 \\

2001-05-29 & $0.9^{+0.4}_{-0.3}$ & $1.5 \times 10^{38}$ & $6.5^{+0.9}_{-0.5}
\times 10^{22}$ & & $58^{+\infty}_{-41}$ & $7.3 \times 10^{40}$ & 106 / 93 \\

\enddata

\tablenotetext{a}{Model = [Galactic absorption] $\times$ ( [bremsstrahlung] +
  [intrinsic absorption] $\times$ \{ [power law] \emph{or} [bremsstrahlung] \}
  ). The subscript $l$ refers to parameters of the low energy component,
  modeled as bremsstrahlung emission.  The subscript $h$ refers to the
  parameters of the high energy component, modeled as either a power-law or
  bremsstrahlung emission.}

\tablenotetext{b}{0.5--2~keV luminosity of the low-energy bremsstrahlung
  component, corrected for Galactic absorption.}

\tablenotetext{c}{Intrinsic equivalent hydrogen column density.}

\tablenotetext{d}{2--10~keV luminosity of the high-energy power-law or
  bremsstrahlung component, corrected for intrinsic and Galactic absorption.}

\tablenotetext{e}{Results corrected for the effects of pile-up.}

\label{tab:spec:continuum}
\end{deluxetable}

%%%%%%%%%%%%%%%%%%%%%%%%%%%%%%%%%%%%%%%%%%%%%%%%%%%%%%%%%%%%%%%%%%%%%%

\vfil\eject\clearpage
\begin{deluxetable}{ccccc}
\tablewidth{0pt} %\rotate
  
\tablecaption{Iron Line Parameters for the Nucleus of NGC 4258}

\tablecolumns{5}

\tablehead{
  \colhead{Date of obs.} &
  \colhead{Line energy (keV)\tablenotemark{a}} &
  \multicolumn{3}{c}{Equivalent Width (eV)\tablenotemark{b}} \\
  
  \colhead{} &
  \colhead{90\%\ conf.} &
  \colhead{90\%\ conf.} &
  \colhead{99\%\ conf.} &
  \colhead{99.5\%\ conf.}
}

\startdata

2000-03-08 & 6.40\tablenotemark{c} & $-240 < {\rm EW} < 887$ & \\

2000-04-17 & $6.41^{+0.05}_{-0.06}$ & $-166^{+62}_{-100}$ &
$-166^{+108}_{-110}$ & $-166^{+115}_{-138}$ \\

2000-04-17 & $6.87^{+0.09}_{-0.06}$ & $-215^{+115}_{-123}$ &
$-215^{+182}_{-186}$ & $-215^{+199}_{-203}$ \\

2000-04-17 (0 -- 6 ksec)\tablenotemark{d} & $6.36^{+0.15}_{-0.13}$ &
$-105^{+104}_{-79}$ & & \\

2000-04-17 (0 -- 6 ksec)\tablenotemark{d} & $6.95^{+0.22}_{-0.12}$ &
$-450^{+172}_{-280}$ & $-450^{+280}_{-396}$ & $-450^{+306}_{-413}$ \\

2000-04-17 (6 -- 12 ksec)\tablenotemark{d} & $6.44^{+0.10}_{-0.16}$ &
$-249^{+147}_{-205}$ & $-249^{+162}_{-290}$ & $-249^{+209}_{-293}$ \\

2000-04-17 (6 -- 12 ksec)\tablenotemark{d} & $6.97$\tablenotemark{c} & $-208 <
{\rm EW} < 323$ & &  \\

2001-05-28 & 6.40\tablenotemark{c} & $-41 < {\rm EW} <132$ & \\

2001-05-29 & 6.40\tablenotemark{c} & $-72 < {\rm EW} < 94$ & \\

\enddata

\tablenotetext{a}{Measured in the rest frame of NGC 4258.}

\tablenotetext{b}{Positive values of EW indicate emission and negative values
  of EW indicate absorption.}

\tablenotetext{c}{Parameter fixed.}

\tablenotetext{d}{The 2000-04-17 observation has been split into two
  consecutive segments each of 6 ksec duration as indicated in parentheses.}

\label{tab:fe_lines}
\end{deluxetable}

%%%%%%%%%%%%%%%%%%%%%%%%%%%%%%%%%%%%%%%%%%%%%%%%%%%%%%%%%%%%%%%%%%%%%%

\vfil\eject\clearpage

%%%%%%%%%%%%%%%%%%%%%%%%%%%%%%%%%%%%%%%%%%%%%%%%%%%%%%%%%%%%%%%%%%%%%%
 
\vfil\eject\clearpage

%%%%%%%%%%%%%%%%%%%%%%%%%%%%%%%%%%%%%%%%%%%%%%%%%%%%%%%%%%%%%%%%%%%%%%

\vfil\eject\clearpage

%%%%%%%%%%%%%%%%%%%%%%%%%%%%%%%%%%%%%%%%%%%%%%%%%%%%%%%%%%%%%%%%%%%%%%

\vfil\eject\clearpage

%%%%%%%%%%%%%%%%%%%%%%%%%%%%%%%%%%%%%%%%%%%%%%%%%%%%%%%%%%%%%%%%%%%%%%

\vfil\eject\clearpage

%%%%%%%%%%%%%%%%%%%%%%%%%%%%%%%%%%%%%%%%%%%%%%%%%%%%%%%%%%%%%%%%%%%%%%

\vfil\eject\clearpage

%%%%%%%%%%%%%%%%%%%%%%%%%%%%%%%%%%%%%%%%%%%%%%%%%%%%%%%%%%%%%%%%%%%%%%

\end{document}